\documentclass[conference]{IEEEtran}
\IEEEoverridecommandlockouts
\usepackage{cite}
\usepackage{amsmath,amssymb,amsfonts}
\usepackage{algorithmic}
\usepackage{graphicx}
\usepackage{textcomp}
\usepackage{xcolor}
\usepackage{booktabs}
\usepackage{multirow}

\usepackage{fancyhdr}    
\def\BibTeX{{\rm B\kern-.05em{\sc i\kern-.025em b}\kern-.08em
    T\kern-.1667em\lower.7ex\hbox{E}\kern-.125emX}}
\begin{document}

\title{An Efficient FPGA Accelerator for Point Cloud}

\author{\IEEEauthorblockN{Zilun Wang, Wendong Mao, Peixiang Yang, Zhongfeng Wang, and Jun Lin }\\
\IEEEauthorblockA{School of Electronic Science and Engineering, Nanjing University, P. R. China\\
Email: \{zlwang,wdmao,pxyang\}@smail.nju.edu.cn, \{zfwang,jlin\}@nju.edu.cn}}

\maketitle

\begin{abstract}
Deep learning-based point cloud processing plays an important role in various vision tasks, such as autonomous driving, virtual reality (VR), and augmented reality (AR). The submanifold sparse convolutional network (SSCN) has been widely used for the point cloud due to its unique advantages in terms of visual results. However, existing convolutional neural network accelerators suffer from non-trivial performance degradation when employed to accelerate SSCN because of the extreme and unstructured sparsity, and the complex computational dependency between the sparsity of the central activation and the neighborhood ones. In this paper, we propose a high performance FPGA-based accelerator for SSCN. Firstly, we develop a zero removing strategy to remove the coarse-grained redundant regions, thus significantly improving computational efficiency. Secondly, we propose a concise encoding scheme to obtain the matching information for efficient point-wise multiplications. Thirdly, we develop a sparse data matching unit and a computing core based on the proposed encoding scheme, which can convert the irregular sparse operations into regular multiply-accumulate operations. Finally, an efficient hardware architecture for the submanifold sparse convolutional layer is developed and implemented on the Xilinx ZCU102 field-programmable gate array board, where the 3D submanifold sparse U-Net is taken as the benchmark. The experimental results demonstrate that our design drastically improves computational efficiency, and can dramatically improve the power efficiency by 51 times compared to GPU.

\end{abstract}

\section{Introduction}\label{sec:intro}
Three dimensions (3D) point cloud is the inherently sparse data acquired from 3D sensors and can provide rich geometric, shape, and scale information~\cite{survey}. Compared with two dimensions (2D) RGB images, 3D point cloud preserves a better understanding of the original geometric information in 3D space for deep learning-based vision tasks. While the biggest challenge of computing on the 3D point cloud comes from its extremely sparse nature. What’s more, the sparsity of point cloud is fundamentally different from that in traditional convolutional neural networks (CNNs). For CNNs, the sparsity is usually caused by the activation functions. But for point cloud, its sparsity reflects the 3D composition of the real world. How to reduce the redundant computation caused by high sparsity 
\let\thefootnote\relax\footnotetext{
This work was supported in part by the National Natural Science Foundation of China under Grant 62174084, 62104097 and in part by the High-Level Personnel Project of Jiangsu Province under Grant JSSCBS20210034, the Key Research Plan of Jiangsu Province of China under Grant BE2019003-4. (Corresponding author: Zhongfeng Wang; Jun Lin.)}
\begin{figure}[hbt]
\centering
    \includegraphics[width=3.5in]{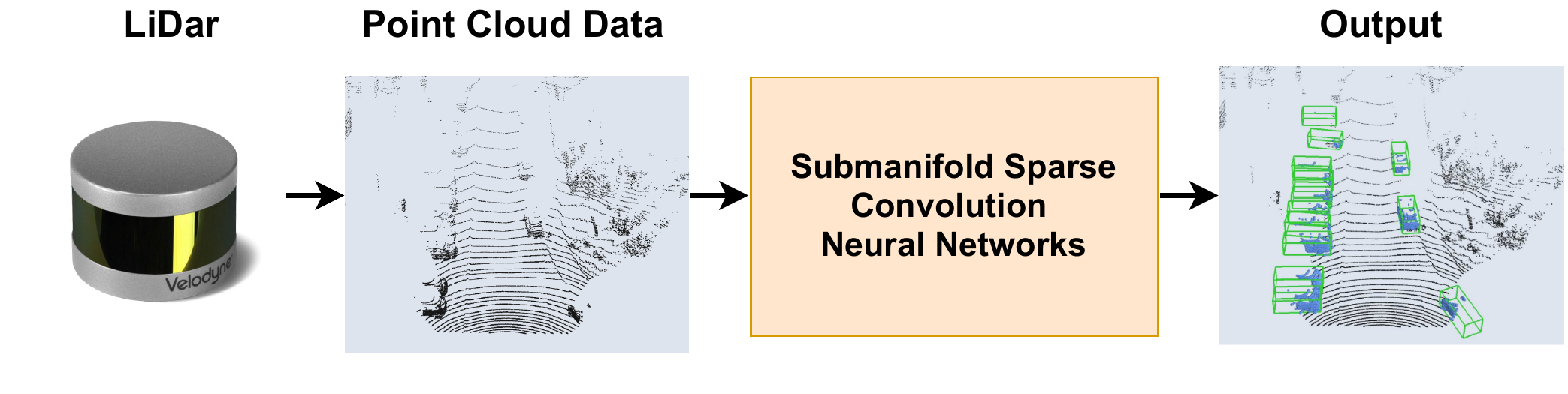}
    \caption{An example of the point cloud application~\cite{pointapp}.}
\label{fig:pointclouds}
\end{figure}
becomes the key to the processing of point cloud. Prior works have proposed deep learning-based methods. For instance,~\cite{1pointcloud2d,2pointcloud2d,3pointcloud2d} projected the 3D point cloud into 2D to compress the data dimensions and reduce computational complexity, then applied 2D CNNs on the 2D point cloud.~\cite{1pointbased,2pointbased,3pointbased} directly leveraged the Multi-Layer Perceptrons (MLPs) operation for the original points to extract semantic features from the sparse point cloud, without voxelizing point cloud into 3D grids.~\cite{1volxel,2volxel,3volxel,ssc} converted the point cloud into sparse discrete representation, then applied modified 3D CNNs for different tasks. Furthermore, authors in~\cite{ssc} proposed submanifold sparse convolution (Sub-Conv) to reduce memory and computational costs of computing on the point cloud by restricting the computation of convolution to be related to nonzero activations. The submanifold sparse convolutional network (SSCN)~\cite{ssc} achieves remarkable results compared to other deep learning-based methods~\cite{facebook}. Consequently, SSCN plays an important role in point cloud-based deep learning applications, motivating its deployment on resource constraint edge devices and corresponding dedicated accelerators.

Nowadays, to accelerate CNNs, some specifically designed hardware accelerators~\cite{1cnnacc,2spareacc,GoSPA} are proposed. Eyeriss~\cite{1cnnacc} presented a general dataflow to minimize data movement. GoSPA~\cite{GoSPA} proposed an intersection method to optimize the dataflow when the activations and weights had sparsity. However, when these accelerators for CNNs are directly used for SSCN, they suffer from severe performance degradation because they can not perform the matching operation of explicitly determining each nonzero activation and searching its nonzero neighbors, which is the core operation of the Sub-Conv layer. Therefore, a dedicated accelerator for SSCN is highly desired to promote its deployment. 

Currently, several works presented solutions for point cloud-based networks.~\cite{pointnetacc} and~\cite{mesorasi} introduced ASIC-based accelerators for PonitNet++ and proposed optimization schemes to the neighbor point search.~\cite{FPGA} designed a low-power FPGA-based accelerator, which optimized the nonlinear implementations in PointNet. The above works are based on the PointNet and PointNet++ networks, and thus cannot be directly applied to the acceleration of SSCN. PointAcc~\cite{pointacc} proposed an ASIC-based accelerator that unified diverse mapping operations into a multiply-accumulate operation through coordinate transformation to be compatible with different point cloud networks. Other hardware solutions such as GPUs can be deployed to accelerate the point cloud networks. However, GPUs are not suitable for resource constraint edge devices because of their high power consumption, and the matching operation also limits their performance. Concentrating on the SSCN,  we propose an FPGA-based efficient SSCN accelerator, ESCA, to support the matching operation and corresponding computations. This work makes the following contributions: 
\begin{itemize}
\item A tile-based zero removing strategy is proposed to improve computational efficiency. The strategy reduces the processing time of the sparse information significantly, which also alleviates the computational load imbalance.
\item An encoding scheme is introduced to efficiently support the matching operation. Based on the above scheme, a matching method is proposed to execute the matching operation for each nonzero activation, which solves the problems of explicit representation in the matching operation.
\item A dedicated SSCN accelerator is proposed to support the matching operation and corresponding computations. The proposed design is implemented in the Xilinx ZCU102 platform and achieves significant improvement in terms of GOPS and power efficiency compared with GPU.
\end{itemize}
\section{Background}\label{sec:bg}
The computation rules of Sub-Conv are fundamentally different from that of traditional convolution. Fig. 2(a) shows the results of traditional convolution for sparse features, and Fig. 2(b) shows the matching process of Sub-Conv. In traditional convolution, the input feature map is traversed by a kernel, and multiply-accumulate operations are performed in order. Even if the feature map has sparsity, as long as the convolution parameters, such as stride, kernel size, etc., are determined, the computation rules and correspondences in the convolution are explicitly determined. As a result, the sparse data in the output feature map dilates~\cite{ssc}, so it is not suitable for point cloud-based computation.

For Sub-Conv~\cite{ssc}, the fields of the feature map involved in the convolution operations are strictly limited to the neighbors of the nonzero activations, and the output feature map maintains the same sparsity as the input feature map. As shown in Fig. 2(b), five nonzero activations mean that this feature map only needs to perform five convolution operations with the corresponding kernel, and the positions are strictly limited to the fields where the central activation is nonzero. Because the Sub-Conv layer can keep the same pattern of sparsity between the input feature map and the output feature map, it shows satisfying visual results when is applied to the point cloud with high sparsity.

However, because the restricted computation pattern of the Sub-Conv layer leads to irregular sparse matching 
\begin{figure}[hbt]
\centering
    \includegraphics[width=3.3in]{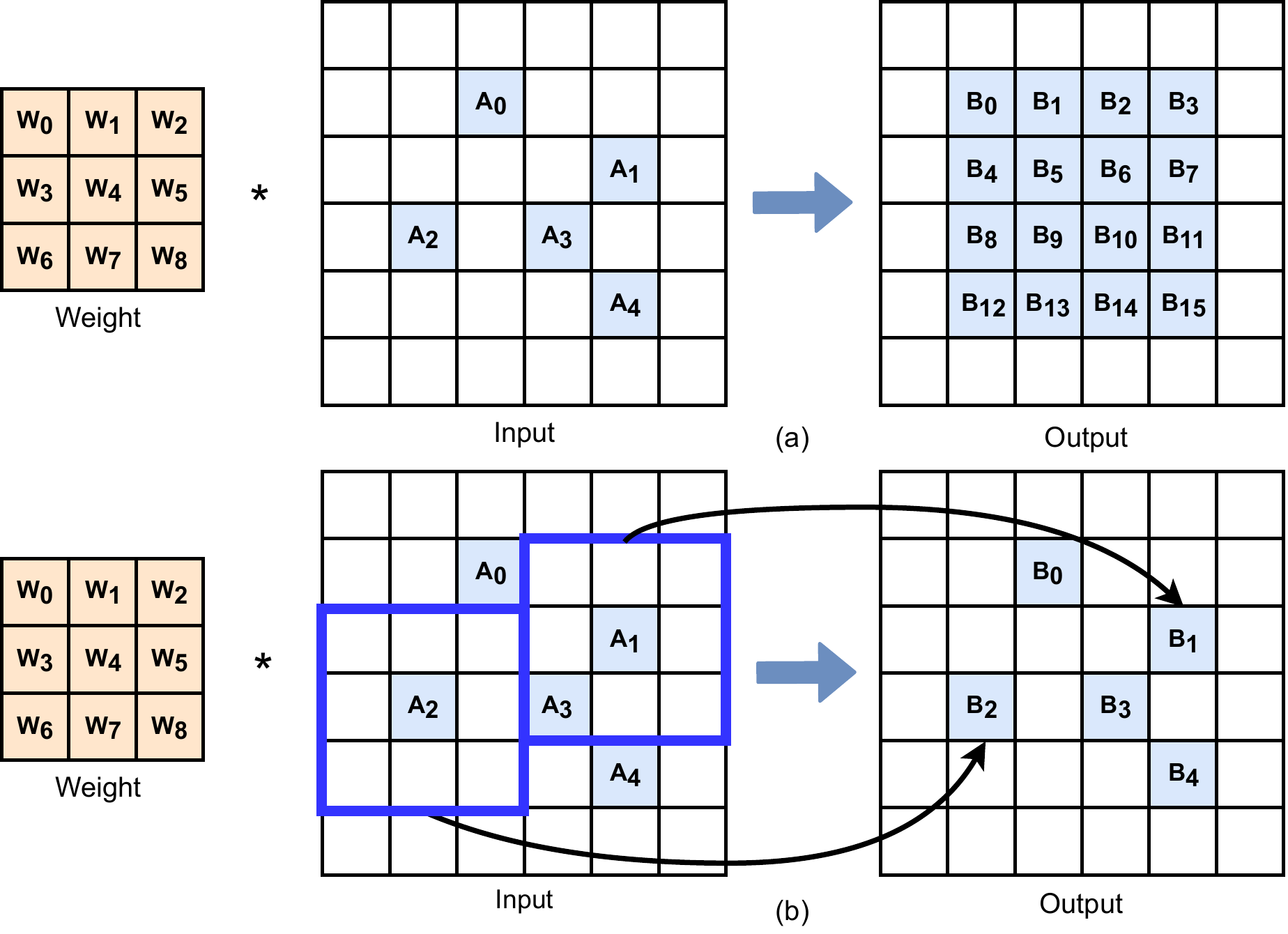}
    \caption{Illustration of traditional convolution and Sub-Conv. (a) Traditional convolution: the feature map is traversed by the kernel, and the sparsity in the output feature map dilates.  (b) Sub-Conv: The kernel only calculates with the fields where the center activation is non-zero.
}
\label{fig:spareconv.pdf}
\end{figure}
operations, traditional convolution accelerators suffer from performance degradation when they are directly applied to it~\cite{pointacc}. Therefore, efficient accelerators for SSCN are urgently needed, and the bottleneck lies in the extreme and unstructured sparsity, and the complex computational dependency between the sparsity of the central activation and the neighborhood ones.
\section{Efficient Design For Submanifold Sparse Convolutional Network}\label{sec:m}

\subsection{Tile-based Zero Removing Strategy}\label{ssec:m}
Voxelized point cloud has huge sparsity. Directly processing on the original feature map results in large memory overhead and computation cost, and dramatically reduce computational efficiency. Take the ShapeNet dataset~\cite{shapenet} as an example, it has nearly 99.9\% sparsity, resulting in many regions without nonzero activations. Since the computation depends on the sparsity of the central activation, removing the all-zero regions has no effect on the result. To tackle this problem, we propose an effective tile-based zero removing strategy to remove the coarse-grained redundant sparse regions.
\begin{figure}[hbt]
\centering
    \includegraphics[width=3.5in]{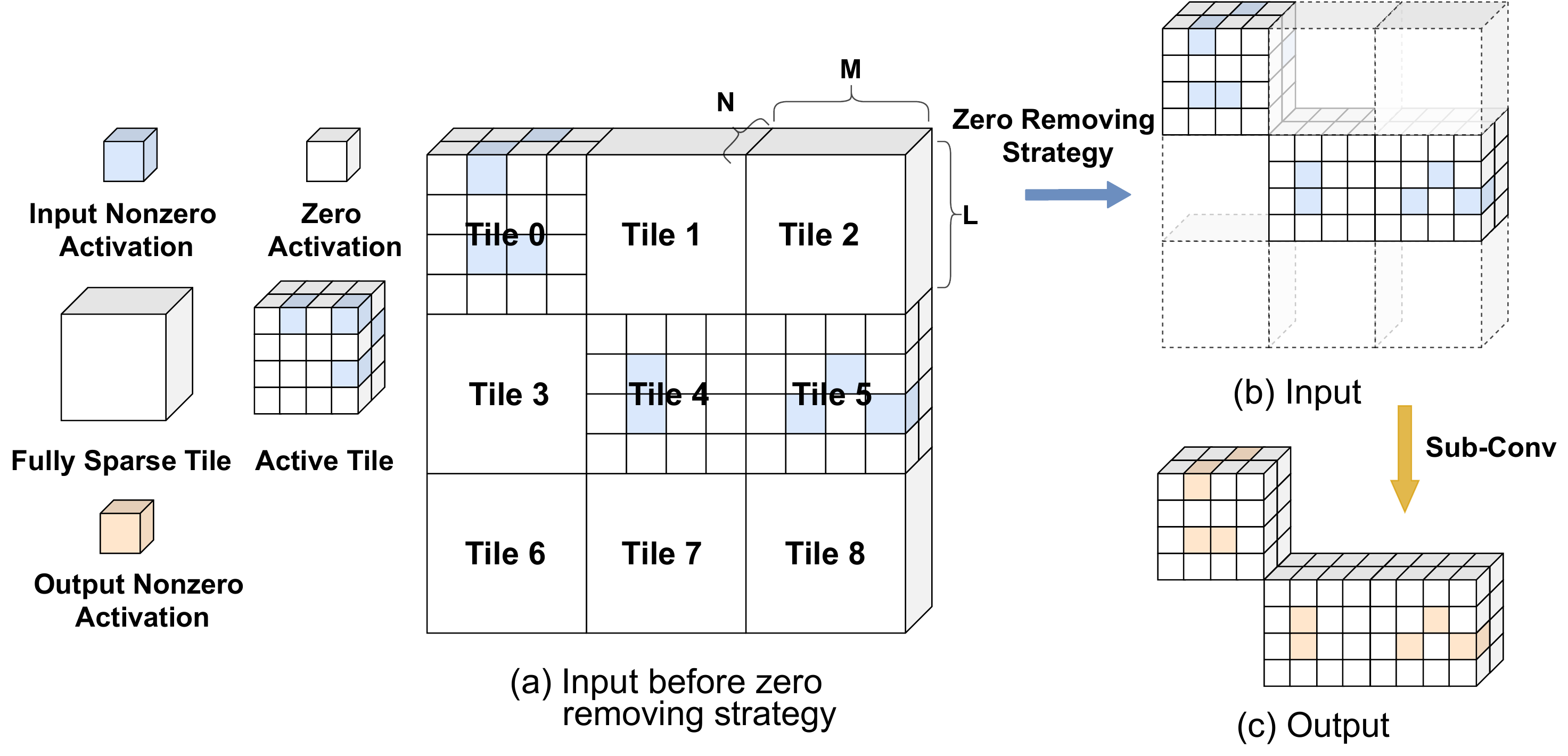}
    \caption{The process of zero removing strategy. (a) The original input feature map is first divided into tiles of fixed size. (b) The fully sparse tiles of the input are removed, keeping only tiles containing nonzero activations. (c) Due to the nature of Sub-Conv, the removal of fully sparse tiles does not affect the output.}
\label{fig:tile}
\end{figure}
As illustrated in Fig. 3(a), the original 3D feature map is divided into tiles of size ${N \times M \times L}$, where ${N}$, ${M}$ and ${L}$ are configurable parameters, and the sparsity in each tile is detected. If all the activations are zero in the tile, the tile is fully sparse and will be removed from the original feature map as shown in Fig. 3(b). Because the fully sparse tile is irrelevant to the computation of the Sub-Conv, the output feature map, as depicted in Fig. 3(c), still maintains the same sparsity. Then the processed feature map is only composed of active tiles, which contain at least one nonzero activation, and will be sequentially matched and computed. With this zero removing strategy, the time overhead when processing sparse information is significantly reduced, and the problem of computational imbalance is also alleviated.

\subsection{Matching Operation and Encoding Scheme}\label{ssec:m}
Matching operation is the procedure to locate each nonzero activation and search its nonzero neighbors, which is crucial for the computation of SSCN, and the position information recording the geometric distribution of nonzero activations is required to support the matching operation. Thus, an encoding scheme is proposed, which encodes the feature map into two types of data: index mask and valid data.

\textbf{Index Mask}. The index mask is used to explicitly represent the sparsity distributions of the feature map and is dynamically traversed during computation. The relationship between features, masks, and nonzero activations is shown in Fig. 4. Mask is a one-bit signal with only two states of 0 and 1, which represents that the activation is zero or not, respectively, and it is stored in the mask buffer. It also has a strong correlation with the sparse distribution of the point cloud, so the computation relationship between input feature maps and matching operation can be established explicitly.

\textbf{Valid Data}. Valid data are the nonzero activations and the corresponding weights, as shown in Fig. 4. As valid data, the activations and weights are stored in the corresponding buffers, and can be read from the buffers under the guidance of the index mask. Thus, the matching operation can be performed through the process of interaction between the index mask and the valid data.
\begin{figure}[hbt]
\centering
    \includegraphics[width=3.5in]{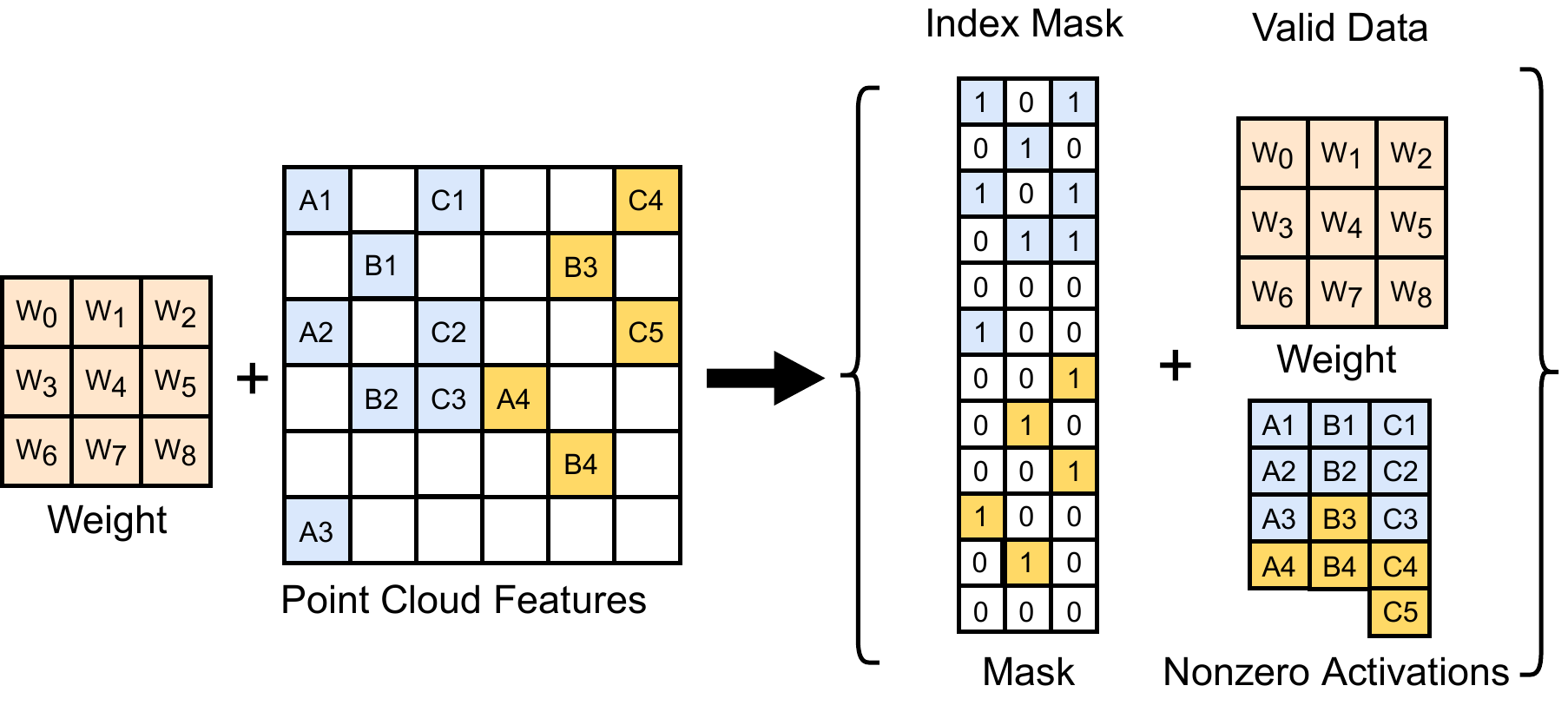}
    \caption{Composition of the index mask and the vaild data.}
\label{fig:mask.pdf}
\end{figure}

\begin{figure}[hbt]
\centering
    \includegraphics[width=3.5in]{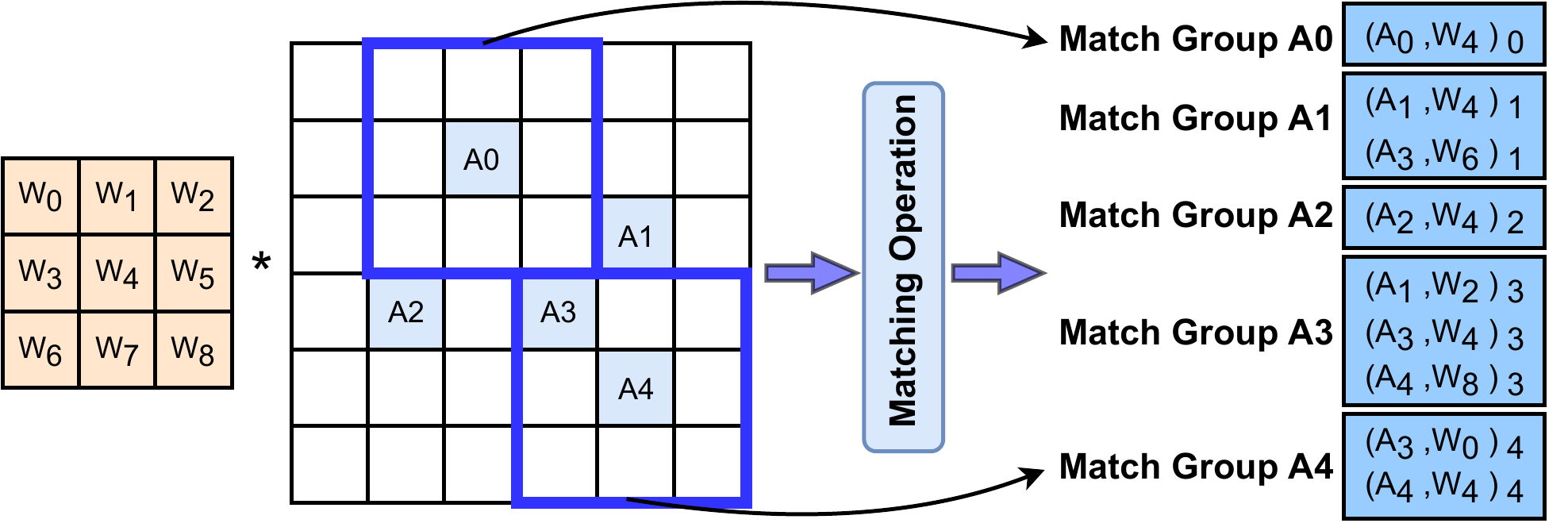}
    \caption{Illustration of the matching operation and match group.}
\label{fig:matchgroup}
\end{figure}
\begin{figure}[hbt]
\centering
    \includegraphics[width=3.5in]{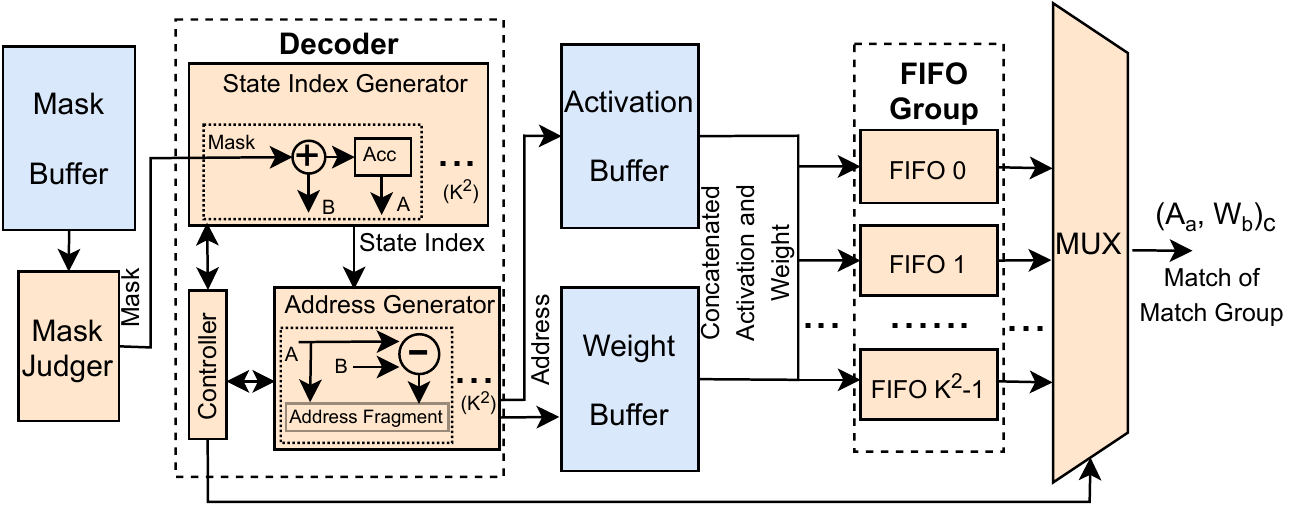}
    \caption{Description of the SDMU. The Acc in the state index generator corresponds to the accumulation operation.}
\label{fig:sparse unit.pdf}
\end{figure}
\subsection{Sparse Data Matching Unit}\label{ssec:arc}
The matching operation and the composition of match group are elaborated in Fig. 5. A \textbf{match group}
contains the nonzero activations and corresponding weights for each convolution calculation based on the central nonzero activation. Also a set of elements in a match group is called a match. Thus, after determining all the match groups for each nonzero activation, the matching operation is completed for one feature map. Meanwhile, the computation of the Sub-Conv layer is decomposed into point-wise multiply-accumulate operations for each match group.
 
To support the matching operation and search all match groups efficiently for the Sub-Conv layer, we propose the sparse data matching unit (SDMU), which is shown in Fig. 6. The mask judger and the decoder perform the matching operation and generate the match groups from the buffers. For the convolution with the kernel size of ${K \times K \times K}$, the index masks of each column are read sequentially. So the parallelism of the decoder in SDMU is ${K^2}$, which corresponds to the number of columns. Then the FIFO group stores the match groups in column order. Finally, the multiplexer (MUX) selects matches from the FIFO group and sends them to the computing core for point-wise multiply-accumulation.


To coordinate the computation rules, activations and the ones that are in their neighbor field need to be explicitly acquired at the same time. Therefore, ${K^compute array3}$ masks are required for determination. This area is called the sparse receptive field (SRF). For each nonzero activation, the matching operation and the acquisition of the match group are limited to the SRF.

The process of matching operation is described in Fig. 7(a).  In this case, it is presented in 2D and can be smoothly extended to 3D. The kernel size is ${3^2}$, so the parallelism is ${3}$. The following steps of the matching method, read masks,
\begin{figure}[hbt]
\centering
    \includegraphics[width=3.5in]{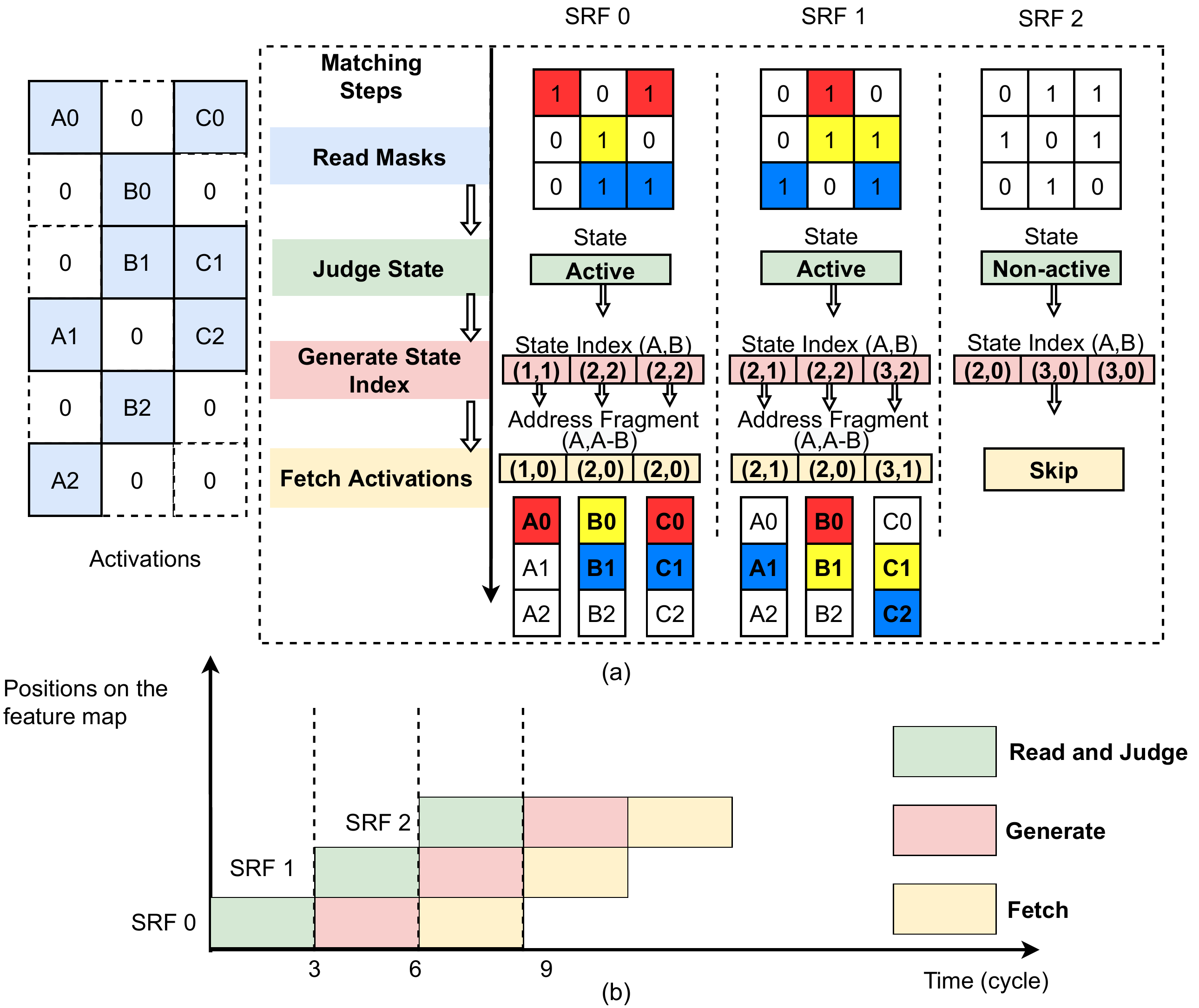}
    \caption{Examples of the matching steps in the SDMU. (a) The process of obtaining match groups through masks.  (b) Pipeline representation when executing the matching operation.}
\label{fig:statevalue.pdf}
\end{figure}
judge state, generate state index, and fetch activations are presented to conduct the matching operation.

\textbf{Read masks:} The index mask is read from the mask buffer for each SRF and sent to the mask judger.

\textbf{Judge state:} The mask is judged whether to perform the convolution for the SRF by the mask judger. If the center mask corresponds to a nonzero activation, then this SRF is active, and the match group is fetched from buffers according to the generate state index step and fetch activations step. Otherwise, it is non-active and the fetch activations step will be skipped.
 
\textbf{Generate state index:} In this step, the relative position of nonzero activations is generated for each SRF and is called the state index. It can be regarded as an array \emph{(A, B)}. The index \emph{A} records the nonzero activations accumulated in each column and it is cumulated for each SRF. The index \emph{B} represents the number of activations in each column for each SRF if the state is active, otherwise, index \emph{B} equals 0. Thus, the index \emph{A} marks the highest address of the activation in the activation buffer for each match group. And the index \emph{B} corresponds to the address length of the activation involved in the computation in each column. 


\textbf{Fetch activations:} The address fragment for nonzero activations of each column can be represented by \emph{(A, A-B)}. It is generated in the address generator and contains addresses for all activations in each match group. Then the corresponding activations are read from the activation buffer. If the mask of the central site is zero, which indicates the matching operation will not be implemented, the fetch activations step for this SRF will be skipped accordingly.

These matching steps are executed in a pipeline, as shown in Fig. 7(b). Since weights and activations have a positional correspondence in each match group, the weights that need to participate in the computation can also be obtained by state index synchronously, and the corresponding activations and weights are concatenated when read from buffers. In summary, the state index obtained by traversing the index mask can establish a matching relationship with valid data, through which the match group can be collected.

In the matching steps, parallel processing is performed according to the column dimension in every SFR to maintain the synchronization of explicit representations of each match group. Therefore, after obtaining the match group from ${K^2}$ columns, which is decided by the kernel size, a FIFO group is applied to store them. The FIFO group consists of ${K^2}$ identical FIFOs, and each FIFO stores the matches belonging to one column. In each cycle, the controller in the decoder selects a match from a FIFO based on the calculation order, and MUX sends it to the computing core.
\subsection{Computing Core}\label{ssec:arc}
Since the sparse data are already transformed into match groups in the SDMU, the computing core (CC) is designed to implement dense point-wise multiply-accumulate operations. The CC contains a computing array and an accumulator. In each cycle, the input to the computing array is a match belonging to a match group. In order to improve throughput, the computing array is divided into $m+1$ computing units (CUs), each of which performs the computation of $n+1$ input channels (ICs), and the output of each CU is the partial sum of the corresponding output channel (OC), so the total parallelism of the computing array is $(m+1)×(n+1)$. 

Fig. 8(b) illustrates the inputs and outputs of the computing array. The activations of the $n+1$ ICs are broadcast to all CUs. \emph{$A_{[n]}$} represents activations belonging  to IC \emph{$n$}. \emph{$W_{[n][m]}$} represents weights belonging to IC \emph{$n$}, OC \emph{$m$}.
For example, the result of CU $m$ is equal to the partial sum of the $n$ ICs on the $m^{th}$ OC.

The detailed structure of the computing unit is shown in Fig. 8(c). The partial sum of nonzero activations for different OCs can be obtained through the computing array, then the partial sum is sent to the accumulator and the output of each SRF is obtained. 

\begin{figure}[hbt]
\centering
    \includegraphics[width=3.5in]{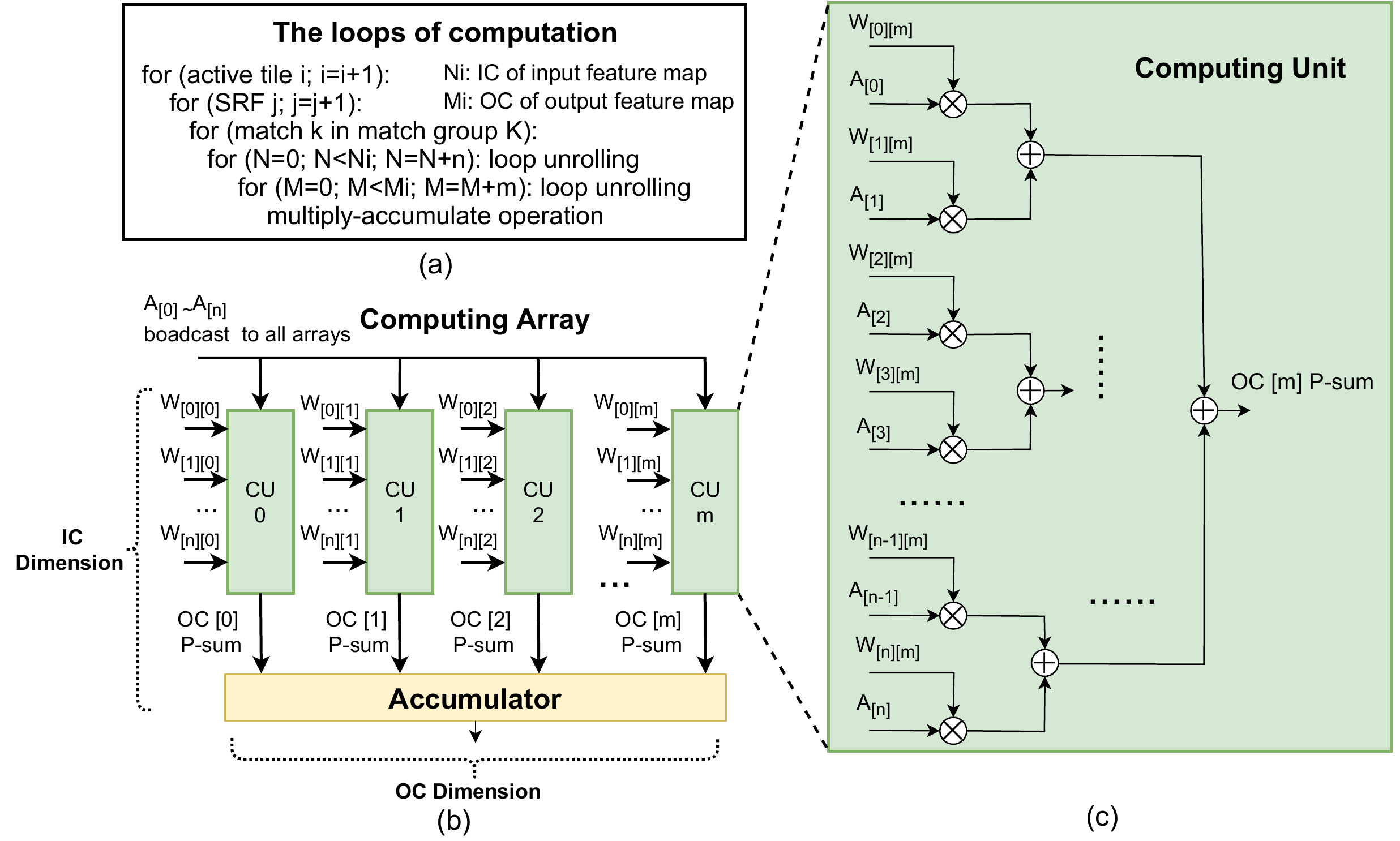}
    \caption{Illustration of loop unrolling and the composition of computing array. (a) The process of loop unrolling. (b) The description of the computing array. (c) The structure of the computing unit in the computing array.}
\label{fig:compute array.pdf}
\end{figure}
\begin{figure}[hbt]
\centering
    \includegraphics[width=3.5in]{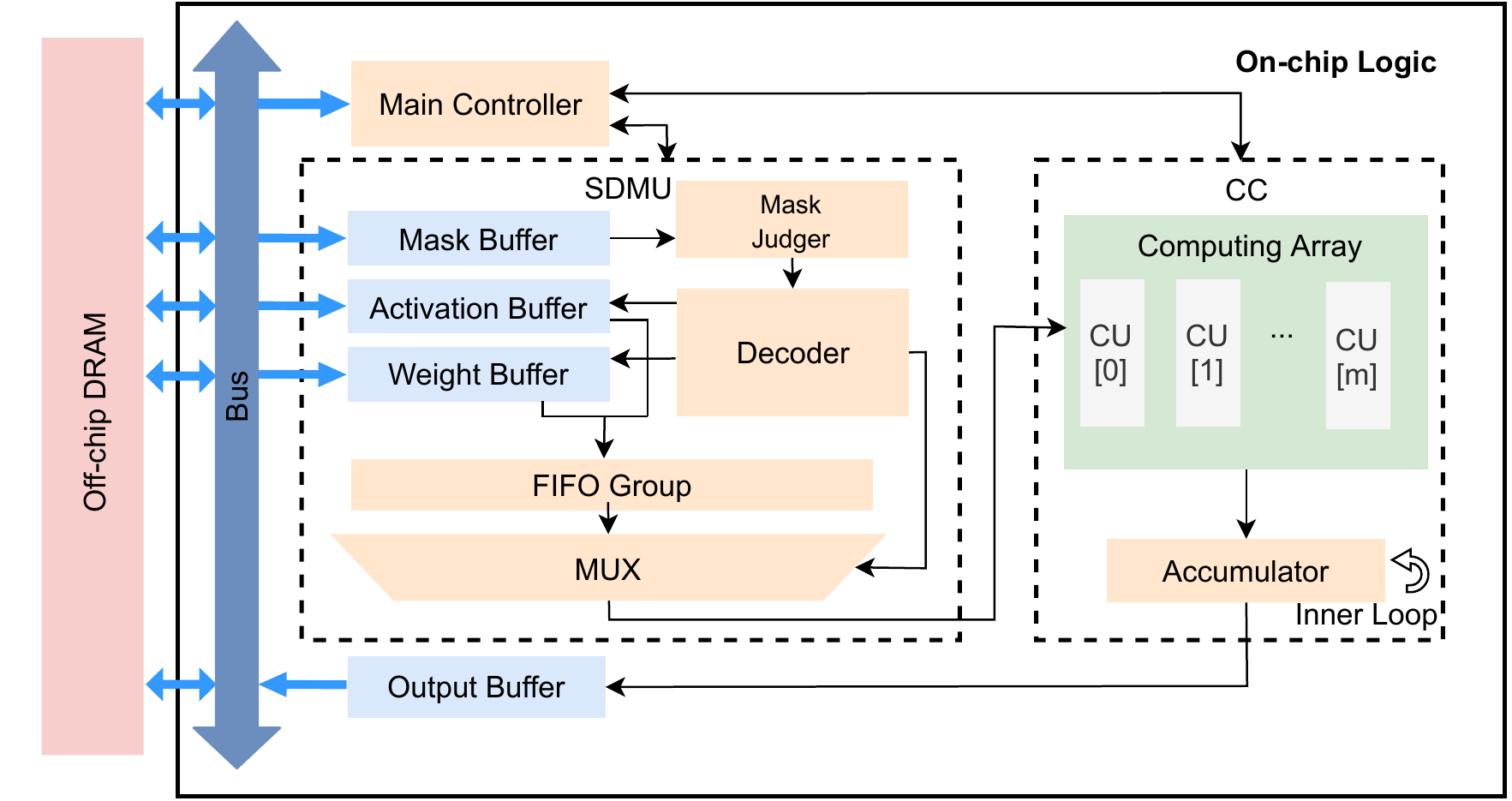}
    \caption{Description of overall hardware architecture.}
\label{fig:overview.pdf}
\end{figure}
The details of the loops are shown in Fig. 8(a). Each active tile is traversed in turn. The obtained data are fed to the CC in the order of matched nonzero activations and weights, and the IC and OC dimensions are completed sequentially according to the parallelism of the proposed computing array. Finally, the partial sum of each match group is accumulated to obtain the outputs corresponding to nonzero activations. The SDMU and CC are executed in pipeline to increase resource utilization and the system throughput.

\subsection{Overall Hardware Architecture}\label{ssec:arc}
The overall hardware architecture is shown in Fig. 9, mainly containing a main controller, an SDMU, a CC, and corresponding buffers on the on-chip logic. 

\textbf{Main Controller}. The main controller is responsible for ensuring that the SDMU and the CC are executed in the right order. 


\textbf{SDMU}. In SDMU, the mask judger and the decoder perform the matching operation. The obtained match groups are stored in the corresponding FIFOs, so as to read them under the control of the FIFO group and MUX, and the matched data are sent to the computing array in order.


\textbf{CC}. In the computing array of CC, computation is performed in the IC and OC dimensions, and the partial sum is generated in the OC dimension. Then the partial sum is accumulated in the accumulator and finally sent to the output buffer. In our structure, the parallelism is set to 16 both in the OC and IC dimensions.

There are four \textbf{buffers} to store data, whose basic unit is block RAM. The mask buffer stores the mask, while the activation buffer and weight buffer store activations and weights respectively. The output buffer stores the outputs and sends them to the off-chip DRAM.
\section{Experimental Results} \label{sec:results}

\subsection{Experimental Setup}\label{ssec:results}
We adopt the 3D submanifold sparse U-Net (SS U-Net)~\cite{ssc} to evaluate our ESCA. SS U-Net can perform the semantic segmentation task of the point cloud with satisfactory visual results. The pre-trained network parameters are 8bit quantized, and the activations are 16bit quantized. The kernel size of the Sub-Conv in the SS U-Net is ${3\times3\times3}$, so the parallelism of SDMU and the number of FIFOs in the FIFO group are set to ${3^2}$. The whole system is implemented with Vivado Design Suite. The performance of the GPU baseline is measured by NVIDIA System Management Interface.

\subsection{Analysis of Zero Removing Strategy}\label{ssec:results}
We comprehensively evaluate the zero removing strategy on two representative point cloud datasets, ShapeNet dataset~\cite{shapenet} and NYU Depth dataset (v2)~\cite{nyu}. The feature maps are normalized to the size of ${192 \times 192 \times 192}$ after voxelization. We test the effect of different tiling sizes on the sparsity and the number of remaining active tiles. The experimental results are shown in Table \uppercase\expandafter{\romannumeral1}. With different tiling sizes, this strategy achieves up to 99.82\% zero reduction in the ShapeNet~\cite{shapenet}, and up to 99.85\% in the NYU~\cite{nyu}. A more fine-grained tile size increases the removing ratio of zeros, it also increases the computational complexity. We use the tile size of ${8 \times 8 \times 8}$.


\begin{table}[hbt]
\caption{Analysis of Zero Removing Strategy}\label{tab:comp}
\centering
\begin{tabular}{|c|c|c|c|c|}
\hline
\multirow{5}{*}{\begin{tabular}[c]{@{}c@{}}\textbf{ShapeNet}\\ ~\cite{shapenet}\end{tabular}} & \textbf{Tile Size}     & \begin{tabular}[c]{@{}c@{}}\textbf{Active}\\ \textbf{Tiles}\end{tabular} & \begin{tabular}[c]{@{}c@{}}\textbf{All}\\ \textbf{Tiles}\end{tabular} & \begin{tabular}[c]{@{}c@{}}\textbf{Removing}\\ \textbf{Ratio}\end{tabular} \\ \cline{2-5} 
                                                                             & ${4 \times 4 \times 4}$    & 198         & 110592 & 99.82\%      \\ \cline{2-5} 
                                                                             & ${8 \times 8 \times 8}$    & 42          & 13824 & 99.69\%      \\ \cline{2-5} 
                                                                             & ${12 \times 12 \times 12}$ & 23          & 4096 & 99.43\%      \\ \cline{2-5} 
                                                                             & ${16 \times 16 \times 16}$ & 14          & 1728 & 99.18\%      \\ \hline
\multirow{5}{*}{\begin{tabular}[c]{@{}c@{}}\textbf{NYU}\\ ~\cite{nyu}\end{tabular}}      & \textbf{Tile Size}     & \begin{tabular}[c]{@{}c@{}}\textbf{Active}\\ \textbf{Tiles}\end{tabular} & \begin{tabular}[c]{@{}c@{}}\textbf{All}\\ \textbf{Tiles}\end{tabular} & \begin{tabular}[c]{@{}c@{}}\textbf{Removing}\\ \textbf{Ratio}\end{tabular} \\ \cline{2-5} 
                                                                             & ${4 \times 4 \times 4}$    & 161         & 110592 & 99.85\%      \\ \cline{2-5} 
                                                                             & ${8 \times 8 \times 8}$    & 33          & 13824 & 99.76\%      \\ \cline{2-5} 
                                                                             & ${12 \times 12 \times 12}$ & 19          & 4096 & 99.53\%      \\ \cline{2-5} 
                                                                             & ${16 \times 16 \times 16}$ & 9           & 1728 & 99.48\%      \\ \hline
\end{tabular}
\end{table}


\subsection{Results Comparison}\label{ssec:results}
The proposed ESCA architecture is implemented on the Zynq UltraScale+ ZCU102 FPGA at 270MHz. The hardware resource utilization is reported in Table \uppercase\expandafter{\romannumeral2}. 
\begin{table}[hbt]
\caption{FPGA Frequency and Resource Utilization}\label{tab:comp}
\centering
\begin{tabular}{ccccc}
\bottomrule
\textbf{Frequency (MHz)}  & \textbf{LUT} & \textbf{FF} & \textbf{BRAM} & \textbf{DSP} \\ \hline
 270  &\begin{tabular}[c]{@{}c@{}}17614\\ (6.43\%)\end{tabular}   &\begin{tabular}[c]{@{}c@{}}12142\\ (2.22\%)\end{tabular}  &\begin{tabular}[c]{@{}c@{}}365.5\\ (40.08\%)\end{tabular}    &\begin{tabular}[c]{@{}c@{}}256\\ (10.16\%)\end{tabular}   \\ \bottomrule
\end{tabular}
\end{table}

ESCA is compared with Tesla P100 GPU and Intel Xeon Gold 6148 CPU, which are existing hardware acceleration solutions for SSCN. As shown in Fig. 10, our ESCA outperforms the CPU and GPU implementation by around 8.41 times and 1.89 times in terms of speedup. Since the computation of SSCN depends on the sparsity of the center activation and its neighborhood ones, the GPU and CPU cannot recognize 
\begin{figure}[hbt]
\centering
    \includegraphics[width=3.5in]{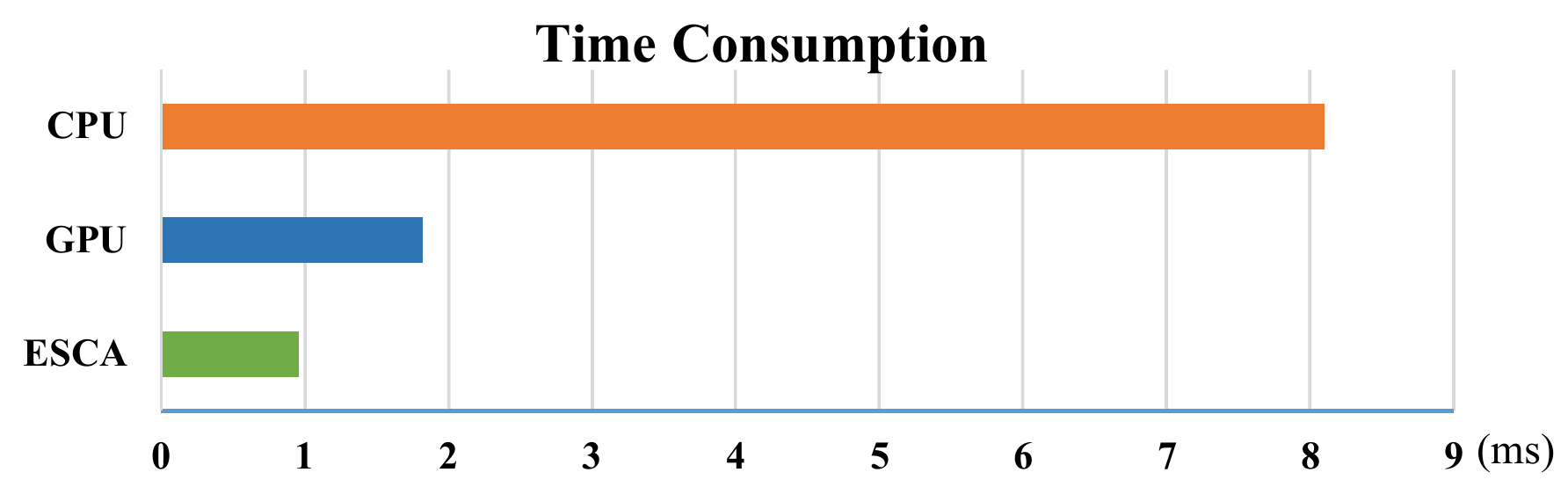}
    \caption{Comparison with CPU and GPU in terms of time consumption when processing a Sub-Conv layer.}
\label{fig:statevalue.pdf}
\end{figure}
\begin{table}[hbt]
\caption{Comparison with Other Implementations for Point Cloud}\label{tab:comp}
\centering
\begin{tabular}{|c|c|c|c|}
\hline                                                                   & \textbf{GPU}        & \textbf{\cite{FPGA}}    & \textbf{ours}       \\ \hline
\textbf{Device}                                                             & Tesla P100 & ZynqXC7z045 & ZynpZCU102 \\ \hline
\textbf{Frequency (MHz)}                                                             & -  & 100 & 270 \\ \hline
\textbf{Model}                                                              & SS U-Net       & O-Pointnet    & SS U-Net       \\ \hline
\textbf{Precision}                                                          & FP32       & INT16       & INT8/INT16 \\ \hline
\textbf{Power (W)}                & 90.56      & 2.15       & 3.45      \\ \hline
\textbf{\begin{tabular}[c]{@{}c@{}}Performance\\ (GOPS)\end{tabular}}       & 9.40       & 1.21        & 17.73      \\ \hline
\textbf{\begin{tabular}[c]{@{}c@{}}Power Efficiency\\ (GOPS/W)\end{tabular}} & 0.10       & 0.56        & 5.14       \\ \hline
\end{tabular}
\end{table} 
this correspondence, resulting in a large number of redundant computations. While in ESCA, the matching operation is executed efficiently. The detailed comparisons between GPU and our design are summarized in Table \uppercase\expandafter{\romannumeral3}. Our design achieves 17.73 GOPS and 5.14 GOPS/W in terms of performance and power efficiency, which outperforms GPU by around 1.88 times and 51 times. Note that the GOPS is effective performance containing only non-zero multiply-accumulate operations for a fair and clear comparison with other implementations. 

To further evaluate the performance of ESCA, it is also compared with an FPGA-based accelerator\cite{FPGA}, which targets the optimized PointNet (O-Pointnet) and leverages the MLP operations for point clouds. Compared with\cite{FPGA}, our accelerator has a significant improvement in both performance and power efficiency as shown in Table \uppercase\expandafter{\romannumeral3}. 

To sum up, the higher performance of ESCA comes from two aspects. On one hand, the zero removing strategy and encoding scheme optimize the data structure to facilitate the match operation. On the other hand, the on-chip logic efficiently performs matching operation and multiply-add computations by the SDMU and CC.

\section{Conclusion} \label{sec: conclusion}
In this paper, we present ESCA, an efficient FPGA-based accelerator that supports SSCN. A zero removing strategy is introduced to remove the coarse-grained redundant regions and an encoding scheme is proposed to simplify the matching operation. Based on the encoding scheme, the sparse data matching unit (SDMU) and the computation core (CC) are developed. The 3D submanifold sparse U-Net is considered for the experiment. The proposed design is implemented on Xilinx ZCU102. The experimental results show that our work outperforms the GPU by around 1.88 times and 51 times in terms of performance and power efficiency.
\bibliographystyle{IEEEtran}
\bibliography{wzl}

\end{document}